\documentclass{elsart}

\usepackage{natbib}
\usepackage {graphicx}
\usepackage{placeins}
\usepackage{longtable,array,tabularx}
\usepackage{amssymb}

\begin{document}

\begin{frontmatter}

% Title, authors and addresses

% use the thanksref command within \title, \author or \address for footnotes;
% use the corauthref command within \author for corresponding author footnotes;
% use the ead command for the email address,
% and the form \ead[url] for the home page:
% \title{Title\thanksref{label1}}
% \thanks[label1]{}
% \author{Name\corauthref{cor1}\thanksref{label2}}
% \ead{email address}
% \ead[url]{home page}
% \thanks[label2]{}
% \corauth[cor1]{}
% \address{Address\thanksref{label3}}
% \thanks[label3]{}

\title{VLT-SINFONI observations of Mrk 609 - A showcase for X-ray active galaxies chosen from a sample of AGN suitable for adaptive optics observations with natural guide stars\thanksref{talk}}
\thanks[talk]{Based on observations with the ESO VLT; 60.A-9041(A).}

% use optional labels to link authors explicitly to addresses:
% \author[label1,label2]{}
% \address[label1]{}
% \address[label2]{}

\author[1]{J. Zuther}
\author[1,2]{J.-U. Pott}
\author[1]{C. Iserlohe}
\author[1]{A. Eckart}
\address[1]{I. Physikalisches Institut, Universit\"at zu K\"oln, Z\"ulpicher Str. 77, D-50937 K\"oln, Germany}
\address[2]{European Southern Observatory (ESO), Garching bei M\"unchen, Germany}
\author[3]{W. Voges}
\address[3]{Max-Planck Institut f\"ur extraterrestrische Physik, Garching bei M\"unchen, Germany}
\begin{abstract}
We will present first results of ESO-VLT AO-assisted integral-field spectroscopy of a sample of X-ray bright AGN with redshifts of $0.04 < z < 1$. We constructed this sample by cross-correlating the SDSS and ROSAT surveys and utilizing typical AO constraints.
%, i.e. guide stars brighter than V<15 and galaxy/guide star separations of less than 40\arcsec. 
This sample allows for a detailed study of the NIR properties of the nuclear and host environments 
%such as colors, star formation, dynamical information, and nuclear excitation,
with high spectral resolution on the 100~pc scale. These objects can then be compared directly to the local ($z<0.01$) galaxy populations (observed without AO) at the same linear scale. As a current example, we will present observations of the $z=0.034$ Seyfert 1.8 galaxy Mrk 609 with the new AO-assisted integral-field spectrometer SINFONI at the VLT. 
%A further example will be NACO imaging of the z=0.15 AGN SDSS J003542.6+004735.1. 
The successful observations show, that in the future - while having observed more objects - we will be able to determine the presence, frequency and importance of nuclear bars and/or circum-nuclear star forming rings in these objects and address the question of how these X-ray luminous AGN and their hosts are linked to optically/UV-bright QSOs, low-z QSOs/radio galaxies, or ULIRGs. 
\end{abstract}

\begin{keyword}
% keywords here, in the form: keyword \sep keyword
galaxies: active \sep  galaxies: fundamental parameters (classification, colors, luminosities, masses, radii, etc.) \sep galaxies: stellar content 
% PACS codes here, in the form: \PACS code \sep code
\PACS 95.85.Jq \sep 98.54.Cm \sep 98.62.Lv
\end{keyword}

\end{frontmatter}

% main text
\section{Introduction}
\label{intro}
A major cornerstone for extragalactic astronomy is the advent of adaptive optics (AO) assisted imaging and spectroscopy on large ground-based telescopes like the Very Large Telescope (VLT), offering a combination of (near) diffraction-limited resolving power and large light-collecting area of 8-10m class telescopes \citep[e.g.][]{brand05}. In addition 3D spectroscopy allows to study both, the morphology and the chemical composition, as well as the dynamics of extragalactic objects at the same time, at an unprecedented depth.

\subsection{Starburst/Seyfert composite galaxies}
\label{sect:composite}
Despite the finding of an apparent coevolution of super massive black holes in the centers of galaxies and their galaxy bulges (hosts) \citep[e.g.][]{page2001} the detailed nature of this interconnection remains mysterious. Is star formation triggered by the active galactic nucleus (AGN) due to radiation pressure and winds from the accretion disk, which disturb the interstellar medium (as discussed by \citet{vanBreugel1993} for 3C~285)? Or can a nuclear starburst component initiate the accretion process onto the black hole \citep{norman1988}? Recently, in their classification study of IRAS selected ROSAT sources, \citet{moran1996} discovered a class of starburst/Seyfert composite galaxies.
They show optical spectra which are dominated by features of starburst galaxies,  using the line diagnostics of \citet{veilleux1987}. Their X-ray luminosity, however, is typical for Seyfert 2 galaxies. A closer look at the spectra reveals some Seyfert-like features, e.g. [OIII]$\lambda 5007$ is significantly broader than all the narrow emission lines in the optical spectrum and in some cases there is a weak broad H$\alpha$ component. There appears a resemblance with narrow-line X-ray galaxies \citep[e.g.][]{boyle1995}, which also show spectra of composite nature. It is still not clear how their strong X-ray emission can be reconciled with the weak/absent optical Seyfert characteristics. The faintness of these objects in the X-ray as well as the optical domain did not allow to study them in detail so far.

Near infrared (NIR) studies, especially integral field spectroscopy, provide powerful means to investigate the (circum-) nuclear properties of the above described AGN. Besides the much lower dust extinction there are a number of NIR diagnostic lines (in emission as well as in absorption) to probe the excitation mechanisms and stellar populations in these objects. Among these are hydrogen recombination lines, rotational/vibrational transitions of H$_2$, stellar features like the CO(2-0) and CO(6-3) absorption band heads and forbidden lines like [FeII] and [SiVI] \citep{hill1999,mouri1994,marconi1994}.

\subsection{Mining the sky: A sample of X-ray bright AGN}
Multi wavelength sky surveys like the Sloan Digital Sky Survey (SDSS) and the ROSAT All Sky Survey (RASS) are very comprehensive databases to search in for targets suitable for very sensitive and high-resolution AO-assisted observations in the NIR. 

\begin{figure}[h!]
\begin{center}
\includegraphics[width=13cm]{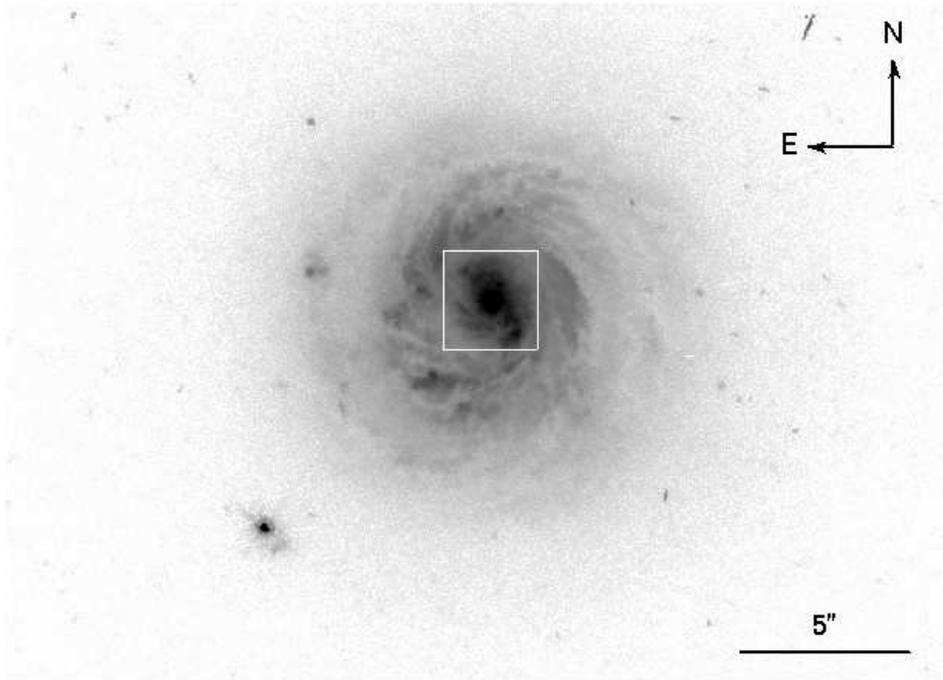}
\caption{HST $V$-band image of Mrk 609 \citep[from the study of][]{malkan1998}. The SINFONI field of view is indicated (see Sect. \ref{sect:sinfo}).}
\label{fig:mrk609_hst}
\end{center}
\end{figure}

In particular, a cross-correlation between the first data release of the SDSS \citep{dr1} and the RASS \citep{voges1999} resulted in a sample of about 70 X-ray luminous AGN ($L_X\approx 10^{43}-10^{45}$~erg~s$^{-1}$) at redshifts between $z=0.1$ and $z=1$ \citep{zu04,zu05}. Most optical counterparts of the X-ray sources turn out to be AGN \citep[e.g.][]{giacconi2001}. Furthermore, these X-ray luminous sources cannot be studied locally because of their small number density in the local universe \citep{hasinger1998}. They are therefore ideal targets for adaptive optics observations. A subset of this sample is comprised of the composite galaxies described in Section \ref{sect:composite}. Our sample allows to study their optical/near-infrared properties in the above described framework. In the following we will present our first integral field observations of the composite galaxy Mrk~609 (Fig. \ref{fig:mrk609_hst}).

\section{SINFONI observations of Mrk~609}
\label{sect:sinfo}
From the subset of starburst/Seyfert composite galaxies we chose Mrk~609 \citep{rudy1988} (Fig. \ref{fig:mrk609_hst}) as one of the closest and brightest (AO self referencing) objects for one of the science verification phase observations\footnote{http://www.eso.org/science/vltsv/sinfonisv/xrayagn.html} of the new AO-assisted integral field spectrometer SINFONI at the Very Large Telescope \citep{sinfoni}. The observations presented in this contribution were taken in AO-mode with a 100~mas pixel scale and a field of view of 3$\times$3 arcsec$^2$. The 2-dimensional image was sliced by small mirrors into 32 slitlets which then were reimaged onto one long pseudoslit and dispersed onto a 2k$\times$2k detector. Using the filters $J$ and $H+K$, a spectral resolution of $R\sim 2000$ was achieved.
Details of the data reduction will be presented in a forthcoming paper (Zuther et al. in prep.). In the following we will present very first results from this study.
\begin{figure}[h!]
\begin{center}
\includegraphics[width=7cm]{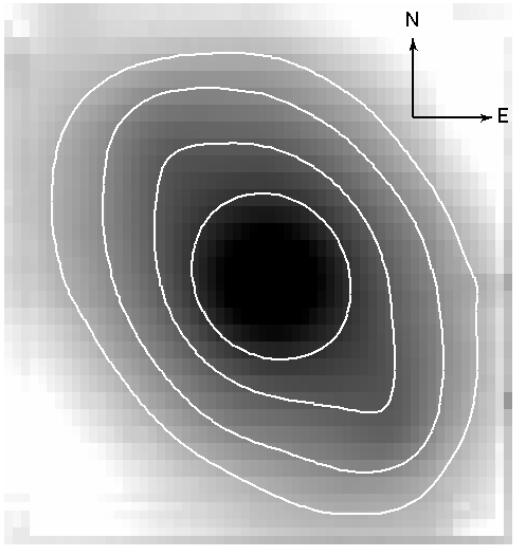}
\caption{$H+K$ median continuum image from the reconstructed data cube. Contours are displayed to guide the eye. The image is $3''\times 3''$ in size.}
\label{fig:hk_median}
\end{center}
\end{figure}

\subsection{Overall properties}
This dataset shows the wealth of information which can be retrieved from the integral field observation. Fig. \ref{fig:hk_median} shows a median $H+K$ continuum image of the reconstructed 3-dimensional data cube. Shown are the inner $3''\times 3''$ around the nucleus. At its redshift of $z=0.034$, 1 arcsecond corresponds to about 700~pc\footnote{Assuming H$_0=70$~km~s$^{-1}$~Mpc$^{-1}$, $\Omega_\Lambda=0$, and $\Omega_m=1$.} and 1 pixel ($\approx$ 0.05 arcsec) to about 40~pc. The shape of the contours, which appear to be elongated towards the root points of the spiral arms (Fig. \ref{fig:mrk609_hst}), suggests the possibility of the presence of a nuclear bar \citep[cf. e.g.][]{martini2001}. Fig. \ref{fig:hk_nuc} shows a nuclear and off-nuclear $H+K$ spectrum. Some properties deserve mentioning:

\begin{figure}
\begin{center}
\includegraphics[width=10cm]{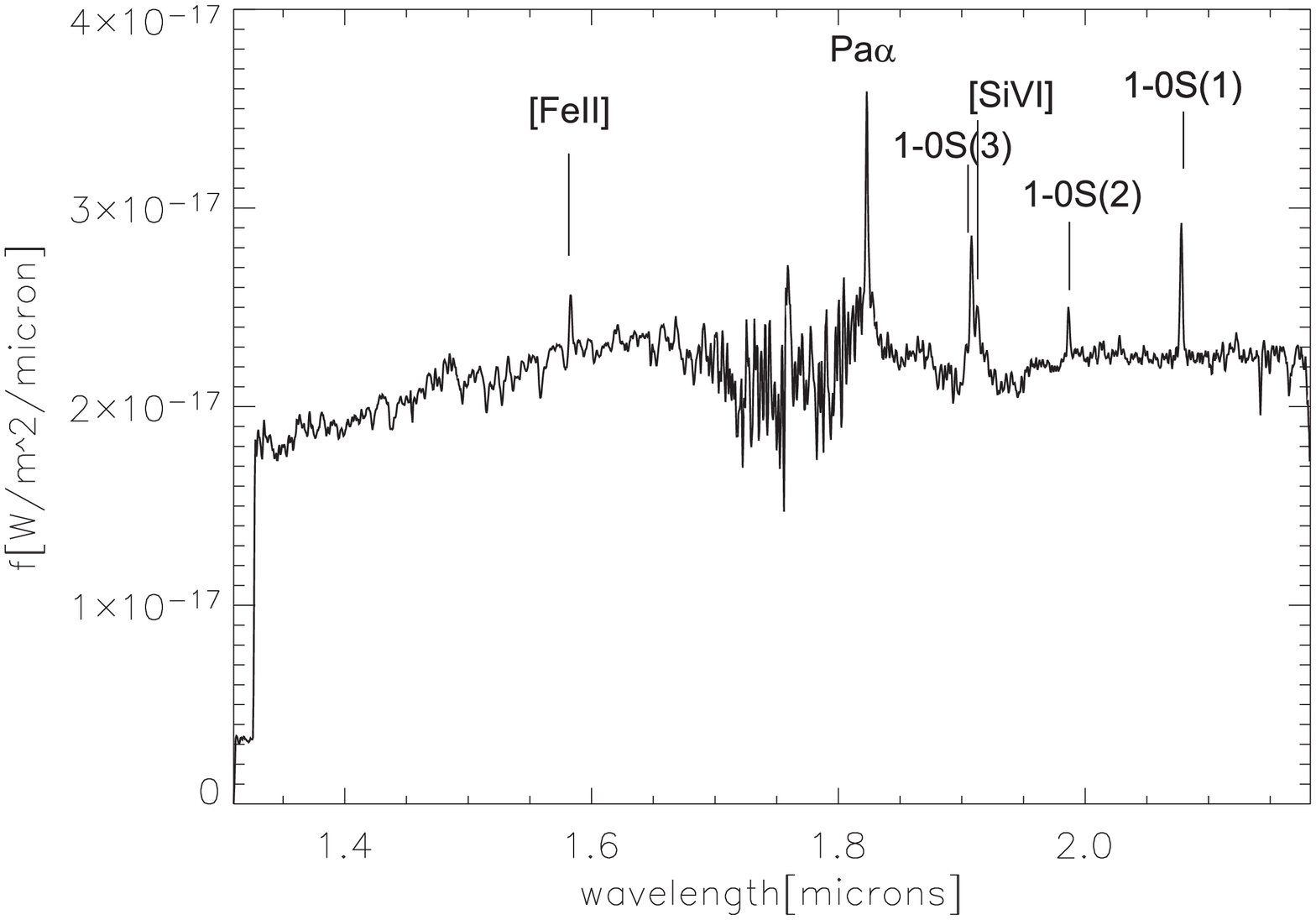}
\includegraphics[width=10cm]{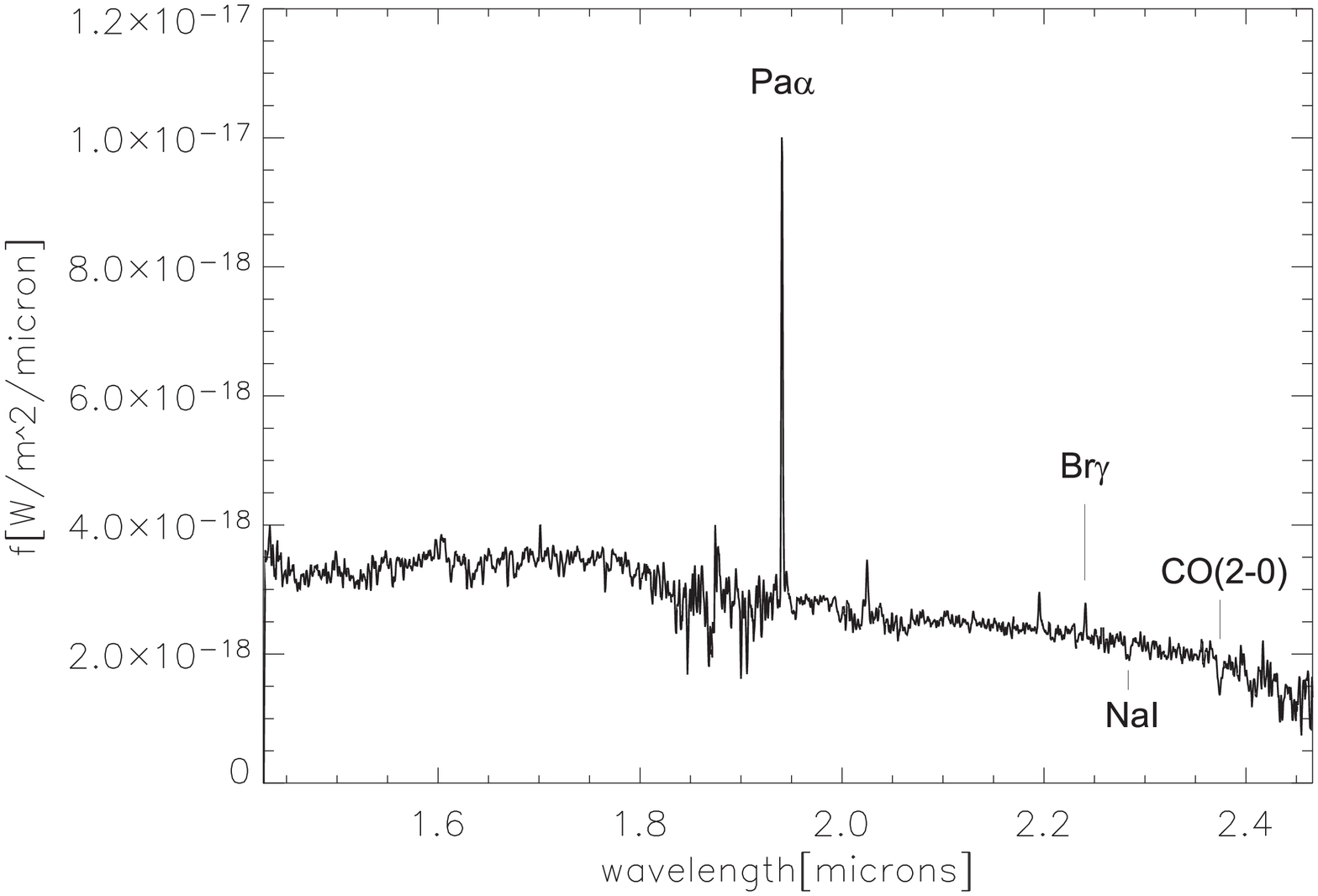}
\caption{The upper spectrum (observed frame) is obtained from the central pixel, the lower spectrum from an off-center position. Line identifications are indicated. Due to a detector defect the nuclear spectrum only extends to 2.2~$\mu$m.}
\label{fig:hk_nuc}
\end{center}
\end{figure}

\begin{itemize}
\item The nuclear spectrum is clearly reddened compared to the off-nuclear one. Reddening towards the nucleus due to the presence of dust is typically found in AGN \citep{glass1985}.
\item The nuclear Pa$\alpha$ line shows a broad component with a width of about 4000~km/s arising from the broad line region. This component is not visible in the off-nuclear spectrum. 
\item Around $1.8\ \mu$m the effect of degrading atmospheric transmission is visible.
\item Stellar absorption features like the NaI~$\lambda 2.206,2.208$ doublet, CaI~$\lambda 2.263$, CO(6-3), and CO(2-0) are visible, helping to estimate the  stellar content.% of the host.
%\item Due to a detector defect, unfortunately, the nuclear spectra are corrupted at 2.3~$\mu$m, the position of the CO(2-0) band head.
\end{itemize}

\subsection{Tracing the narrow-line emitting gas}
\begin{figure}[h!]
\begin{center}
\includegraphics[width=13cm]{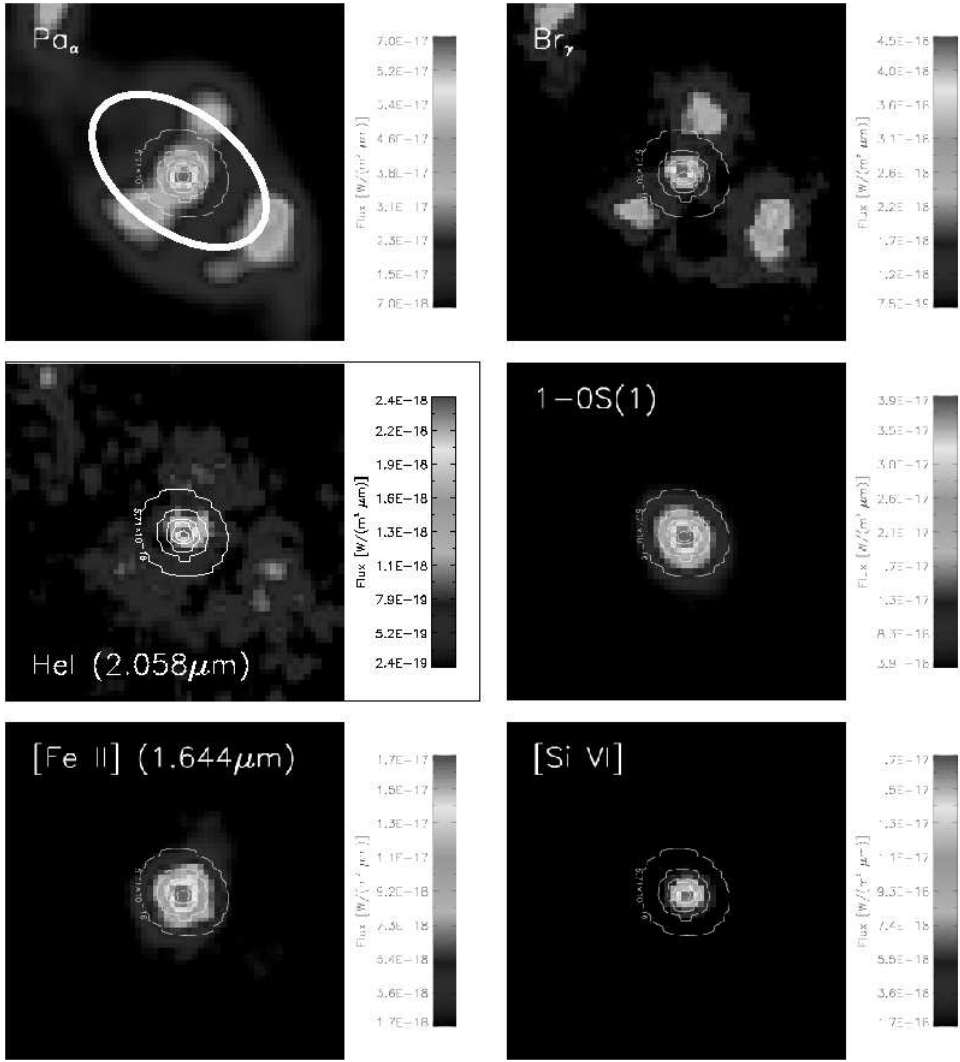}
\caption{Preliminary spatial distribution of prominent emission lines. The potential starburst ring is indicated by a thick ellipse in the Pa$\alpha$ map.}
\label{fig:line_maps}
\end{center}
\end{figure}

The great advantage of integral field spectroscopy is the simultaneous availability of spatial and spectral information. This allows to generate spatial maps of spectral features of interest. The first three panels in Fig. \ref{fig:line_maps} show the recombination lines Pa$\alpha$, Br$\gamma$, and HeI as tracers of star formation. This emission is extended and, besides the nucleus, is concentrated in a ring like structure at a projected distance of about 500~pc. The morphology roughly follows the continuum contours of Fig. \ref{fig:hk_median}. Thus, we could be dealing with a starburst ring embedded in a nuclear bar. There is a plentiful number of examples of starburst rings in galaxies with and without nuclear activity \citep[cf.][and references therein]{smith1999} and which have comparable sizes.

Other species like the forbidden transitions [SiVI] and [FeII] are concentrated on the nucleus in Fig. \ref{fig:line_maps}. High energetic nuclear emission is primarily able to penetrate deeply into the interstellar medium and produce regions which are partly ionized and where such transitions can be excited. The [FeII] emission is slightly extended, indicating also a connection with supernovae excited emission in the starburst ring.

\begin{figure}[h!]
\begin{center}
\includegraphics[width=9cm]{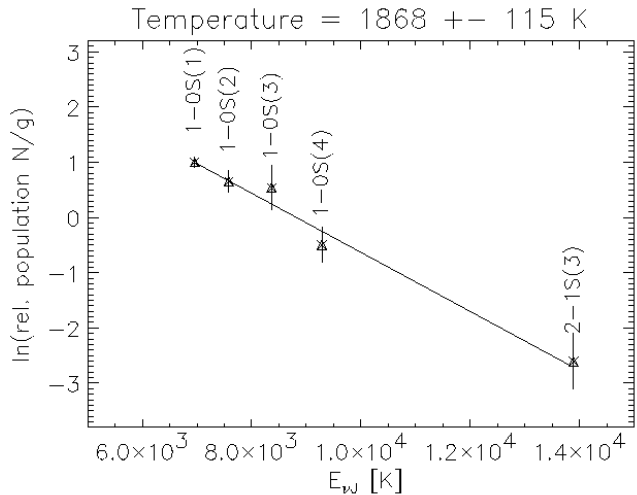}
\caption{H$_2$ level population diagram.}
\label{fig:h2_temp}
\end{center}
\end{figure}

Rotational/vibrational transitions of H$_2$ can be used to study the dominant excitation mechanisms in the circum-nuclear environment \citep[e.g.][]{mouri1994}. The emission lines originate in surfaces of molecular clouds exposed to stellar or nuclear radiation. Three major processes are considered: (1) excitation by X-ray or cosmic rays coming from the AGN; (2) shock excitation either from supernova winds or streaming motion; or (3) UV fluorescence from young stars. The first two processes are of thermal nature, whereas the third is non thermal. 1-0S(1) and the other detectable emission lines are also concentrated on the nucleus. Line ratios are used to estimate the level populations of H$_2$ in Mrk~609, indicating a thermal origin of the emission with a temperature of about 1900~K (Fig. \ref{fig:h2_temp}, Zuther et al. in prep.). At the current stage of analysis, excitation due to X-rays also seems to be unimportant, because a number of other H$_2$ transitions which would be expected in the case of X-ray excitation \citep{tine1997}, are not detected. This was already suggested by previous studies of a number of infrared galaxies \citep[e.g.][]{koorneef1996}.

\begin{figure}[h!]
\begin{center}
\includegraphics[width=9cm]{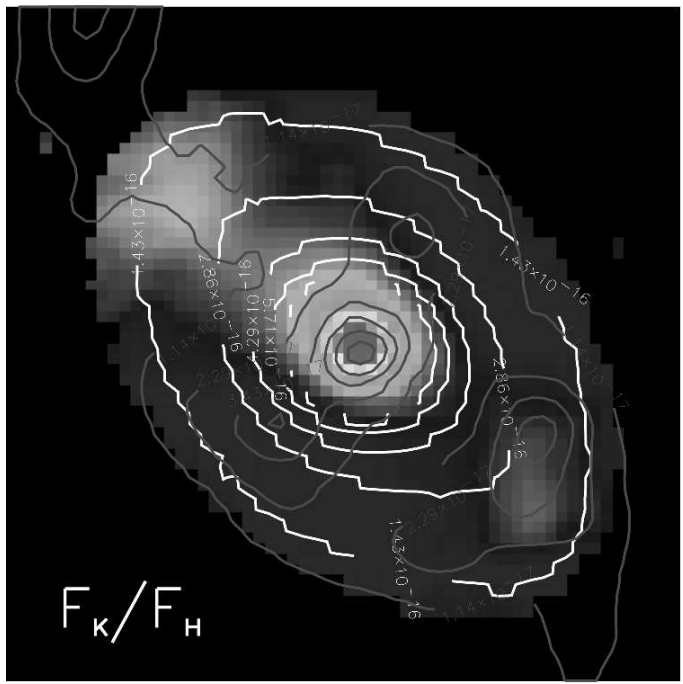}
\caption{Preliminary reddening map. Dark contours correspond to Pa$\alpha$ emission, lighter contours to continuum emission.}
\label{fig:reddening}
\end{center}
\end{figure}

Line ratios of hydrogen recombination lines (e.g. Pa$\alpha$ and Br$\gamma$) can be used to estimate the amount extinction. Assuming case-B recombination we find no significant extinction in the ring like structure. The nuclear ratios also give no significant extinction. At the nucleus, however, this value can be influenced by not-perfect separation of the narrow and broad component of the Pa$\alpha$ line. One can further use parts of the spectrum without prominent emission/absorption lines to generate a reddening map. Fig. \ref{fig:reddening} shows the circum-nuclear reddening using parts of the continuum in the $H$- and the $K$-band. The largest values of reddening are found close to/at the nucleus and at the local Pa$\alpha$ peaks at the tips of the bar within the putative starburst ring. This indicates the presence of dust and molecular material feeding the AGN or the star formation.

\section{Summary and conclusions}
In this contribution we have presented SINFONI science verification observations of the starburst/Seyfert composite galaxy Mrk~609, which has been drawn from a sample of AO-suitable X-ray luminous AGN. AO-assisted integral field spectroscopy in the NIR providing simultaneous spatial and spectroscopic information enables the detailed study of the connection/feedback between star formation and nuclear activity by means of emission/absorption line diagnostics, 2-dimensional morphological, and kinematical analysis. 

The presented observations indicate the presence of a nuclear bar and associated star formation in a starburst ring. The dominating central excitation mechanism for molecular hydrogen appears to be of thermal origin.

Further work on the spectral/spatial analysis will quantify the above statements  and incorporation of complementary multi wavelength data (SDSS, ROSAT, CO(1-0); Zuther et al. in prep.) will give a better understanding of the feedback between nuclear activity and host galaxy environment in Mrk~609. Expanding this study to other members of the sample will provide further insights into the nature of the  class of starbust/Seyfert composites itself.

{\bf Acknowledgments}

This work was supported in part by the Deutsche Forschungsgemeinschaft (DFG) via grant SFB 494.
% The Appendices part is started with the command \appendix;
% appendix sections are then done as normal sections
% \appendix

% \section{}
% \label{}

% Bibliographic references with the natbib package:
% Parenthetical: \citep{Bai92} produces (Bailyn 1992).
% Textual: \citet{Bai95} produces Bailyn et al. (1995).
% An affix and part of a reference:
%   \citep[e.g.][Ch. 2]{Bar76}
%   produces (e.g. Barnes et al. 1976, Ch. 2).

\end{document}